\documentclass[onecolumn,showpacs,preprintnumbers,amsmath,aps]{revtex4}
\usepackage{graphicx}
\usepackage{dcolumn}
\usepackage{bm}
\begin{document}
%%%%%%%%%%%%%%%%%%%%%%%%%%%%
\def\eq#1{(\ref{#1})}
\def\fig#1{\ref{#1}}
\def\tab#1{\ref{#1}}
%%%%%%%%%%%%%%%%%%%%%%%%%%%%
\title{Thermodynamics of the superconducting state in Calcium at 200 GPa}
\author{R. Szcz{\c{e}}{\`s}niak, A.P. Durajski}
%%%%%%%%%%%%
\affiliation{Institute of Physics, Cz{\c{e}}stochowa University of Technology, Al. Armii Krajowej 19, 42-200 Cz{\c{e}}stochowa, Poland}
%%%%%%%%%%%%
\email{adurajski@wip.pcz.pl}
%%%%%%%%%%%%
\date{\today} 
\begin{abstract}
%%%%%%%%%%%%%%%%%%%%%%%%%%%%%%%%%%%%%%%%%%%%%
The thermodynamic parameters of the superconducting state in Calcium under the pressure at 200 GPa were calculated. The Coulomb pseudopotential values ($\mu^{\star}$) from $0.1$ to $0.3$ were taken into consideration. It has been shown, that the specific heat's jump at the critical temperature and the thermodynamic critical field near zero Kelvin strongly decrease with $\mu^{\star}$. The dimensionless ratios $r_{1}\equiv \Delta C\left(T_{C}\right)/C^{N}\left(T_{C}\right)$ and $r_{2}\equiv T_{C}C^{N}\left(T_{C}\right)/H^{2}_{C}\left(0\right)$ significantly differ from the predictions based on the BCS model. In particular, $r_{1}$ decreases from $2.64$ to $1.97$ with the Coulomb pseudopotential; whereas $r_{2}$ increases from $0.140$ to $0.157$. The numerical results have been supplemented by the analytical approach.
%%%%%%%%%%%%%%%%%%%%%%%%%%%%%%%%%%%%%%%%%%%%%
\end{abstract}
\maketitle
%
%%%%%%%%%%%%%%%%%%%%%%%%%%%%%%%%%%%%%%%%%%%%%%%%%%%%%%%%%%%%%%%%%%%%%%%%%%%%%%%%%%%%%%%%%%%%%%%%%%%
\section{Introduction}
%%%%%%%%%%%%%%%%%%%%%%%%%%%%%%%%%%%%%%%%%%%%%%%%%%%%%%%%%%%%%%%%%%%%%%%%%%%%%%%%%%%%%%%%%%%%%%%%%%%

By using the advanced technique, it is experimentally possible to explore the properties of the superconducting state under the high pressure ($p$). In particular, the above researches enable: (i) the test of the theories for the superconducting state, (ii) as well as to improve the properties of the superconductors, and (iii) to create the new superconductors. At present, the $52$ elemental superconductors are known, however $22$ of them superconduct if the pressure is applied \cite{Schilling}. The most interesting elements are: Lithium, where the critical temperature ($T_{C}$) rises rapidly to $\sim 14$ K at $30.2$ GPa \cite{Deemyad}, Yttrium with the maximum value of the critical temperature of $20$ K at $115$ GPa \cite{Hamlin}, and Calcium which has the highest observed value of $T_{C}$ ($25$ K at $161$ GPa) \cite{Yabuuchi}, \cite{Okada}, \cite{Yin}. 

The thermodynamic properties of the superconductors under the high pressure can be analyzed in the framework of the Eliashberg approach \cite{Eliashberg}, \cite{Marsiglio}, \cite{Carbotte}. In this formalism the complicated form of the electron-phonon interaction is modeled by Eliashberg function ($\alpha^{2}F\left(\Omega\right)$). We notice that the first-principle calculations of $\alpha^{2}F\left(\Omega\right)$ require the knowledge of the electronic wave functions, the phonon spectrum, and the electron-phonon matrix elements between two single-electron Bloch states. Experimentally, the form of the Eliashberg function can be directly obtained from the second derivative of $I$-$V$ curve for the tunnel junction ($d^{2}V/dI^{2}\sim \alpha^{2}F\left(mV\right)$) \cite{McMillan}. 

From the physical point of view, the Eliashberg approach represents particularly important method of the analysis, since it enables the calculation of the thermodynamic parameters on the quantitative level. In particular, the exact form of the free energy difference between the superconducting and normal state should be calculated on the basis of the so-called Eliashberg equations \cite{Bardeen}. In the considered case, the input parameters are the Eliashberg function and the Coulomb pseudopotential ($\mu^{\star}$), where $\mu^{\star}$ models the Coulomb repulsion between electrons. We notice that its value is selected in such way that $T_{C}$ determined on the basis of the Eliashberg equations equals the experimental value of the critical temperature.  

In the presented paper, we have analysed the thermodynamic properties of Calcium under the pressure at $200$ GPa by using the Eliashberg approach. In particular, the following parameters were taken into account: the specific heat in the superconducting state ($C^{S}$), the specific heat in the normal state ($C^{N}$), and the thermodynamic critical field ($H_{C}$). Additionally, the dimensionless ratios 
$r_{1}\equiv \left(C^{S}\left(T_{C}\right)-C^{N}\left(T_{C}\right)\right)/C^{N}\left(T_{C}\right)$ and 
$r_{2}\equiv T_{C}C^{N}\left(T_{C}\right)/H^{2}_{C}\left(0\right)$ have been determined \cite{BCS}. 

For Calcium, the dependence of the critical temperature on the pressure has been obtained experimentally by Okada {\it et al.} in 1996 \cite{Okada} and then by Yabuuchi {\it et al.} in 2006 \cite{Yabuuchi}. The results prove that the critical temperature grows with the pressure from the value of about $1$ K ($p = 50$ GPa) to $25$ K ($p=161$ GPa). On the basis of the Yabuuchi's results it is easy to show that the values of the Coulomb pseudopotential can be large. For example, $\left[\mu^{\star}\right]_{{\rm p=120GPa}}=0.215$ and $\left[\mu^{\star}\right]_{{\rm p=160GPa}}=0.241$ \cite{SzczesniakDurajski}, \cite{SzczesniakSzczesniak}. In the case of the pressure $200$ GPa, the experimental value of $T_{C}$ is still unknown, therefore the wide range of the Coulomb pseudopotential's values have been considered in the paper; $\mu^{\star}\in\left<0.1,0.3\right>$.
 
%%%%%%%%%%%%%%%%%%%%%%%%%%%%%%%%%%%%%%%%%%%%%%%%%%%%%%%%%%%%%%%%%%%%%%%%%%%%%%%%%%%%%%%%%%%%%%%%%%%
\section{THE ELIASHBERG EQUATIONS}
%%%%%%%%%%%%%%%%%%%%%%%%%%%%%%%%%%%%%%%%%%%%%%%%%%%%%%%%%%%%%%%%%%%%%%%%%%%%%%%%%%%%%%%%%%%%%%%%%%%

The Eliashberg functions for Calcium were determined by Yin {\it et al.} \cite{Yin}. At the pressure $200$ GPa, the $Pnma$ structure is clearly favored, and the linear-response calculations indicate that it is also dynamically stable. Additionally, the Yin's results have shown that the strong electron-phonon coupling persists and $T_{C}$ can be high ($\sim 30$ K). In the paper, the thermodynamic parameters for Calcium have been calculated by using the Eliashberg equations on the imaginary axis, the Yin's Eliashberg function has been taken into account. 

The Eliashberg set has the following form: 
\begin{widetext}
\begin{equation}
\label{r1}
Z_{n}=1+\frac{1}{\omega_{n}}\frac{\pi}{\beta}\sum_{m=-M}^{M}
                             \lambda\left(i\omega_{n}-i\omega_{m}\right)
                             \frac{\omega_{m}Z_{m}}
                             {\sqrt{\omega_m^2Z^{2}_{m}+\phi^{2}_{m}}},
\end{equation}
and
\begin{equation}
\label{r2}
\phi_{n}=
                                  \frac{\pi}{\beta}\sum_{m=-M}^{M}
                                  \left[\lambda\left(i\omega_{n}-i\omega_{m}\right)-\mu^{\star}\theta\left(\omega_{c}-|\omega_{m}|\right)\right]
                                  \frac{\phi_{m}}
                                  {\sqrt{\omega_m^2Z^{2}_{m}+\phi^{2}_{m}}}.
\end{equation}
\end{widetext}
The wave function renormalization factor is denoted by $Z_{n}\equiv Z\left(i\omega_{n}\right)$, and the order parameter function by $\phi_{n}\equiv\phi\left(i\omega_{n}\right)$. The $n$-th Matsubara frequency has the form: $\omega_{n}\equiv \left(\pi / \beta\right)\left(2n-1\right)$, where $\beta\equiv\left(k_{B}T\right)^{-1}$ and $k_{B}$ is the Boltzmann constant. We notice that the value of the order parameter is given by the ratio: $\Delta_{n}\equiv\phi_{n}/Z_{n}$. In Eqs. \eq{r1} and \eq{r2} the symbol $\lambda\left(z\right)$ denotes the electron-phonon pairing kernel: 
\begin{equation}
\label{r3}
\lambda\left(z\right)\equiv 2\int_0^{\Omega_{\rm{max}}}d\Omega\frac{\Omega}{\Omega ^2-z^{2}}\alpha^{2}F\left(\Omega\right),
\end{equation}
where the maximum phonon frequency $\Omega_{\rm{max}}$ is equal to $78.1$ meV. Finally, $\theta$ denotes the Heaviside unit function and $\omega_{c}$ is the cut-off frequency ($\omega_{c}=3\Omega_{\rm{max}}$). 

From the mathematical point of view the exact solution of the Eliashberg equations represents a complicated problem. We notice that formally the Eliashberg set contains the infinite number of the non-linear algebraic equations; in addition every equation has the integral kernel, which is dependent on the form of the Eliashberg function. It is possible to prove, that if we limit the number of the Matsubara frequencies, the solutions of the Eliashberg equations lose the convergence only in the area of the very low temperatures. In the paper we assume $M=1100$. In this case the functions $Z_{n}$ and $\phi_{n}$ are stable for $T \geq 3.48$ K. The Eliashberg equations have been solved by using iterative method \cite{Szczesniak}.

%%%%%%%%%%%%%%%%%%%%%%%%%%%%%%%%%%%%%%%%%%%%%%%%%%%%%%%%%%%%%%%%%%%%%%%%%%%%%%%%%%%%%%%%%%%%%%%%%%%
\section{THE NUMERICAL RESULTS}
%%%%%%%%%%%%%%%%%%%%%%%%%%%%%%%%%%%%%%%%%%%%%%%%%%%%%%%%%%%%%%%%%%%%%%%%%%%%%%%%%%%%%%%%%%%%%%%%%%%
%

\begin{figure}[ht]%
\includegraphics*[scale=0.40]{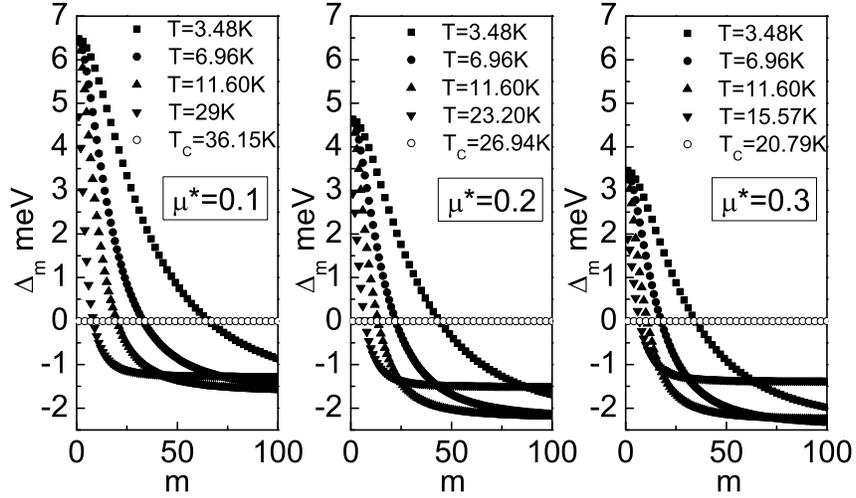}
\caption{The dependence of the order parameter on $m$ for selected values of the Coulomb pseudopotential and temperatures.}
\label{f1}
\end{figure}

\begin{figure}[ht]%
\includegraphics*[scale=0.40]{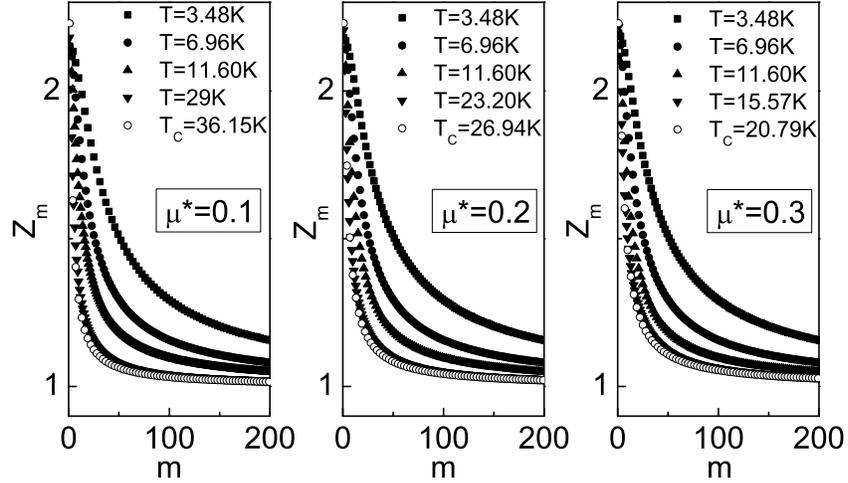}
\caption{The dependence of the wave function renormalization factor on $m$ for selected values of the Coulomb pseudopotential and temperatures.}
\label{f2}
\end{figure}

In Fig. \fig{f1} the order parameter as a function of the number $m$ has been presented. We have considered the selected values of the Coulomb pseudopotential and temperatures. It is easy to see that $\Delta_{m}$ decreases with $m$ number's growth. In particular, the strong fall of the order parameter appears for low values of $m$; for higher values the function $\Delta_{m}$ saturates. Very similar dependence on the number $m$ possesses the wave function renormalization factor (see Fig. \fig{f2}). However, the function $Z_{m}$ saturates considerably slower than $\Delta_{m}$.
     
The growth of the Coulomb pseudopotential's value differently influences on the order parameter and the wave function renormalization factor. In the first case the values of the order parameter strongly decrease; whereas the function $Z_{m}$ is practically not changing. 
From the physical point of view the above facts mean, that together with the increasing of $\mu^{\star}$ decreases only the value of the critical temperature; the electron effective mass remains fixed ($m^{\star}_{e}\sim Z_{m=1}$).

\begin{figure}[ht]%
\includegraphics*[scale=0.40]{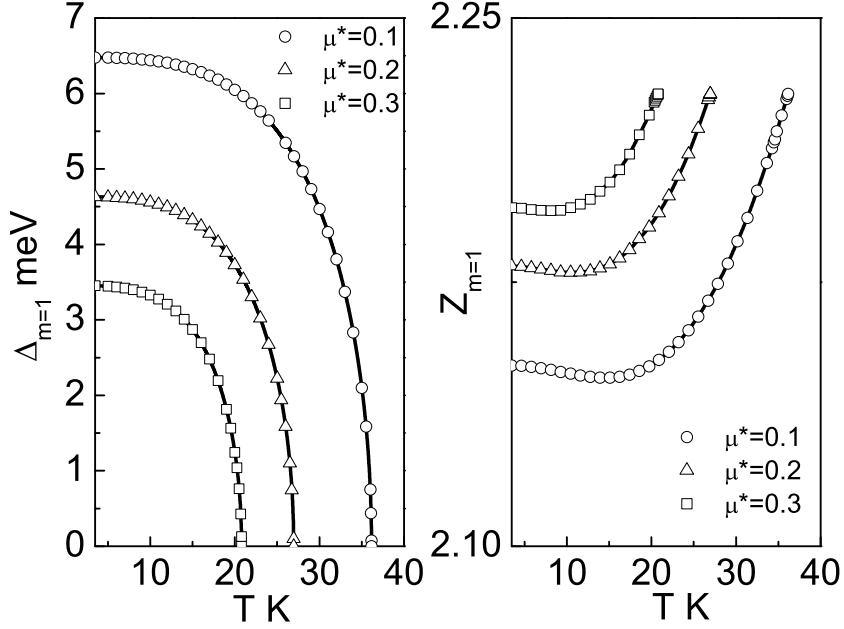}
\caption{The dependence of the order parameter and the wave function renormalization factor on the temperature for selected values of the Coulomb pseudopotential. In both cases we assume $m=1$.}
\label{f3}
\end{figure}

The solutions of the Eliashberg equations also very unlikely evolve with the temperature. Most clearly this fact is possible to observe in Fig. \fig{f3}, where the dependences of $\Delta_{m=1}$ and $Z_{m=1}$ on the temperature have been plotted. In particular, the obtained results show that the temperature dependence of the order parameter can be modeled by the function:
$\Delta_{m=1}\left(T\right)=\Delta_{m=1}\left(0\right)\sqrt{1-\left(\frac{T}{T_{C}}\right)^{\beta}}$, where the parameters $\Delta_{m=1}\left(0\right)$, $T_{C}$, and $\beta$ are collected in Tab. \tab{t1}. On the other hand, the wave function renormalization factor is slightly depended on the temperature. 

%
%%%%%%%%%%%%%%%%%%%%%%%%%%%(rys.1)
%%%%%%%%%%%%%%%%%%%%%%%%%%%
%
%%%%%%%%%%%%%%%%%%%%%%%%%%%(rys.2)
%%%%%%%%%%%%%%%%%%%%%%%%%%%
%

%%%%%%%%%%%%%%%%%%%%%%%%%%%(Tab.1)
\begin{table}
\caption{\label{t1}
The values of $\Delta_{m=1}\left(0\right)$, $T_{C}$, and $\beta$ parameters.}
%\begin{ruledtabular}
\begin{tabular}{ccccccc}
\hline
$\mu^{\star}_{C}$ & & $\Delta_{m=1}\left(0\right)$ meV & &$T_{C}$ K& & $\beta$\\
\hline
$0.1$ & & $6.48$ & &$36.15$& &$3.47$\\
$0.2$ & & $4.63$ & &$26.94$& &$3.50$\\
$0.3$ & & $3.45$ & &$20.79$& &$3.60$\\
\end{tabular}
%\end{ruledtabular}
\end{table}
%%%%%%%%%%%%%%%%%%%%%%%%%%
%
%%%%%%%%%%%%%%%%%%%%%%%%%%%(rys.3)
%%%%%%%%%%%%%%%%%%%%%%%%%%%
%

In order to obtain the specific heats and the thermodynamic critical field, we have to calculate 
the free energy difference between the superconducting and normal state ($\Delta F$) \cite{Bardeen}:
\begin{equation}
\label{r4}
\frac{\Delta F}{\rho\left(0\right)}=-\frac{2\pi}{\beta}\sum_{m=1}^{M}
\left(\sqrt{\omega^{2}_{m}+\Delta^{2}_{m}}- \left|\omega_{m}\right|\right)
(Z^{{\rm S}}_{m}-Z^{N}_{m}\frac{\left|\omega_{m}\right|}
{\sqrt{\omega^{2}_{m}+\Delta^{2}_{m}}}).  
\end{equation}  
The functions $Z^{S}_{m}$ and $Z^{N}_{m}$ represent the wave function renormalization factors for the superconducting (S) and normal (N) state respectively. 

\begin{figure}[ht]%
\includegraphics*[scale=0.40]{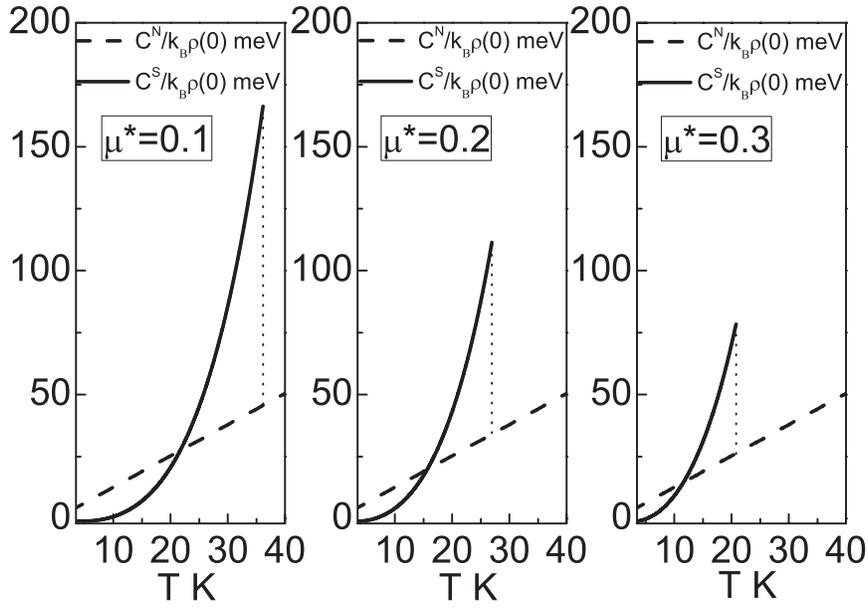}
\caption{The dependence of the specific heat in the superconducting and normal state on the temperature for selected values of the Coulomb pseudopotential. The vertical line indicates the position of specific heat's jump that occurs at $T_{C}$.}
\label{f4}
\end{figure}

The specific heat difference between the superconducting and normal state $\left(\Delta C\equiv C^S-C^N\right)$ can be obtained by using the expression: 
\begin{equation}
\label{r5}
\frac{\Delta C}{k_{B}\rho\left(0\right)}
=-\frac{1}{\beta}\frac{d^{2}\left[\Delta F/\rho\left(0\right)\right]}
{d\left(k_{B}T\right)^{2}}.
\end{equation}
On the other hand, the specific heat in normal state is given as:
\begin{equation}
\label{r6}
\frac{C^{N}}{ k_{B}\rho\left(0\right)}=\frac{\gamma}{\beta}, 
\end{equation}
where $\gamma\equiv\frac{2}{3}\pi^{2}\left(1+\lambda\right)$. In Fig. \ref{f4} the specific heat for the superconducting and normal state as a function of the temperature has been shown. It is easy to see that the specific heat's jump at the critical temperature decreases with the growth of the Coulomb pseudopotential. In particular, 
$\left[\frac{\Delta C\left(T_{C}\right)}{\rho\left(0\right)}\right]_{\mu^{\star}=0.3}/
\left[\frac{\Delta C\left(T_{C}\right)}{\rho\left(0\right)}\right]_{\mu^{\star}=0.1}\simeq 0.43$. 

%
%%%%%%%%%%%%%%%%%%%%%%%%%%%(rys.4)
%%%%%%%%%%%%%%%%%%%%%%%%%%%
%

The thermodynamic critical field has been calculated by using the formula:  
\begin{equation}  
\label{r7}
\frac{H_{C}}{\sqrt{\rho\left(0\right)}}=\sqrt{-8\pi
\left[\Delta F/\rho\left(0\right)\right]}.
\end{equation}

\begin{figure}[ht]%
\includegraphics*[scale=0.40]{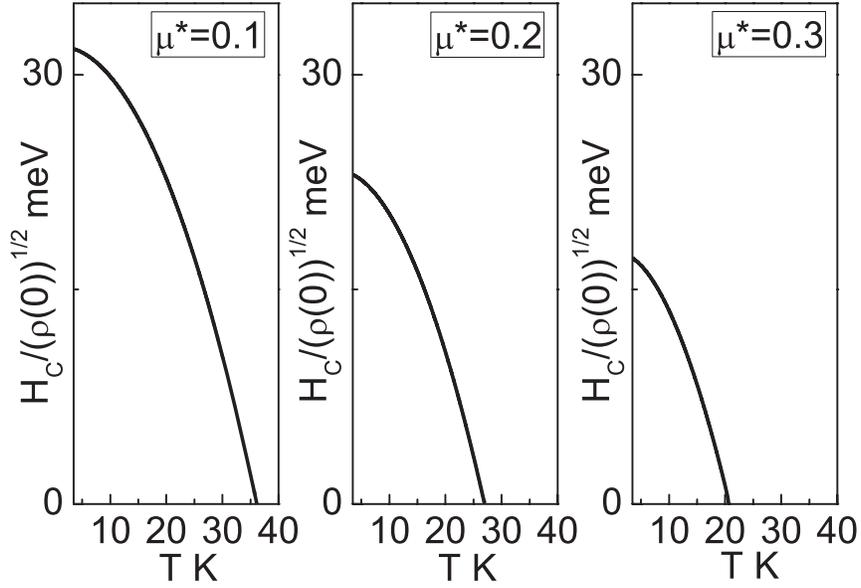}
\caption{
The thermodynamic critical field as a function of the temperature for selected values of the Coulomb pseudopotential.}
\label{f5}
\end{figure}

In Fig. \ref{f5} the dependence of $H_{C}/\sqrt{\rho\left(0\right)}$ on the temperature has been presented. We can see, that the value of the thermodynamic critical field near the temperature of zero Kelvin ($H_{C}\left(0\right)\simeq H_{C}\left(T_{0}\right)$) also strongly decreases with $\mu^{\star}$; 
$\left[\frac{H_{C}\left(0\right)}{\sqrt{\rho\left(0\right)}}\right]_{\mu^{\star}=0.3}/
\left[\frac{H_{C}\left(0\right)}{\sqrt{\rho\left(0\right)}}\right]_{\mu^{\star}=0.1}\simeq 0.55$.

%
%%%%%%%%%%%%%%%%%%%%%%%%%%%(rys.5)
%%%%%%%%%%%%%%%%%%%%%%%%%%%
%

\begin{figure}[ht]%
\includegraphics*[scale=0.40]{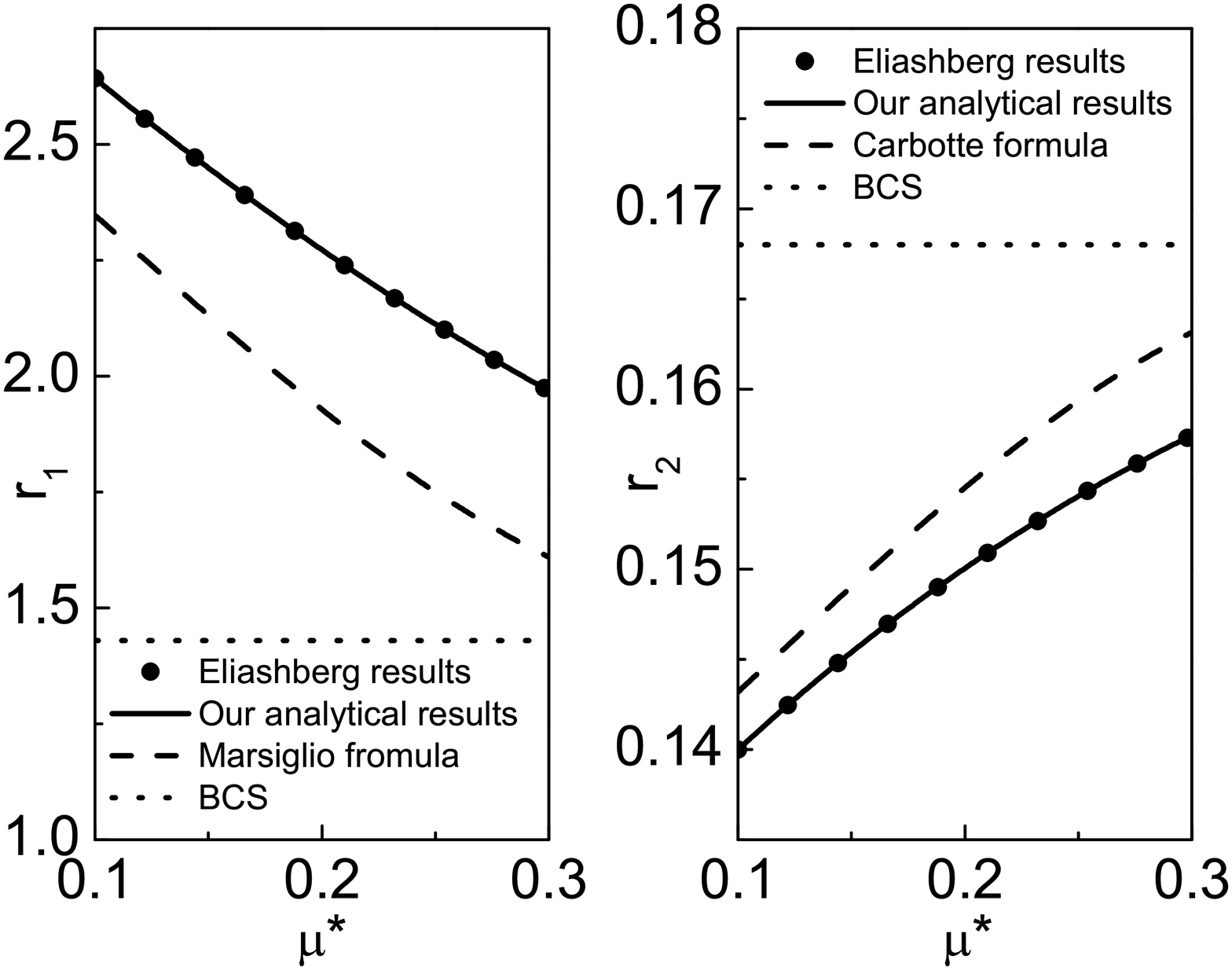}
\caption{The ratios $r_{1}$ and $r_{2}$ as a functions of the Coulomb pseudopotential. The black circles show the exact results obtained on the basis of the Eliashberg equations. The solid lines represent the calculation of the $r_{1}$ and $r_{2}$ parameters using our analytical scheme 
(Eq. \eq{r8} and Eq. \eq{r9}). The dashed lines represent the results obtained in the framework of the Marsiglio \cite{Marsiglio} and Carbotte \cite{Carbotte} formulas respectively. The dotted lines denote the BCS values.}
\label{f6}
\end{figure}

The dimensionless ratios $r_{1}$ and $r_{2}$ on the basis of the calculated thermodynamic functions have been determined. We notice that in the framework of the BCS model these parameters have the universal values: $\left[r_{1}\right]_{\rm BCS}=1.43$ and $\left[r_{2}\right]_{\rm BCS}=0.168$ \cite{BCS}. For Calcium the theoretical data have been collected in Fig. \fig{f6}. We see that even for large values of $\mu^{\star}$ the ratios significantly diverge from the values predicted by the BCS model.
%
%%%%%%%%%%%%%%%%%%%%%%%%%%%(rys.6)
%%%%%%%%%%%%%%%%%%%%%%%%%%%
%
Below, we have given the formulas, which enable the simple calculations of $r_{1}$ and $r_{2}$. In particular:
\begin{equation}
\label{r8}
\frac{r_{1}}{\left[r_{1}\right]_{\rm BCS}}=1-42.5\left[\left(\frac{f_{1}}{f_{2}}\frac{k_{B}T_{C}}{\omega_{\rm ln}}\right)^{2}\ln\left(\frac{3}{2}\frac{f_{1}}{f_{2}}\frac{k_{B}T_{C}}{\omega_{\rm ln}}\right)\right],
\end{equation}
and
\begin{equation}
\label{r9}
\frac{r_{2}}{\left[r_{2}\right]_{\rm BCS}}=1+2.7\left[\left(\frac{f_{1}}{f_{2}}\frac{k_{B}T_{C}}{\omega_{\rm ln}}\right)^{2}\ln\left(\frac{1}{33}\frac{f_{1}}{f_{2}}\frac{k_{B}T_{C}}{\omega_{\rm ln}}\right)\right].
\end{equation}

We notice that the numerical coefficients in Eqs. \eq{r8} and \eq{r9} by the least-squares analysis of $60$ exact $r_{1}$ and $r_{2}$ values have been chosen. Finally, the critical temperature by using the {\it modified} Allen-Dynes formula should be calculated \cite{SzczesniakDurajski1}:
\begin{equation}
\label{r10}
k_{B}T_{C}=f_{1}f_{2}\frac{\omega_{\rm ln}}{1.45}\exp\left[\frac{-1.03\left(1+\lambda\right)}{\lambda-\mu^{\star}\left(1+0.06\lambda\right)}\right],
\end{equation}
where the strong-coupling correction function ($f_{1}$) and the shape correction function ($f_{2}$) are given by:
\begin{equation}
\label{r11}
f_{1}\equiv\left[1+\left(\frac{\lambda}{\Lambda_{1}}\right)^{\frac{3}{2}}\right]^{\frac{1}{3}},
\end{equation}
and
\begin{equation}
\label{r12}
f_{2}\equiv 1+\frac
{\left(\frac{\sqrt{\omega_{2}}}{\omega_{\rm{ln}}}-1\right)\lambda^{2}}
{\lambda^{2}+\Lambda^{2}_{2}}.
\end{equation}
The functions $\Lambda_{1}$ and $\Lambda_{2}$ have the form: 
\begin{equation}
\label{r13}
\Lambda_{1}\equiv 2.6\left(1+1.8\mu^{\star}\right),
\end{equation}
and 
\begin{equation}
\label{r14}
\Lambda_{2}\equiv 0.092\left(1-150\mu^{\star}\right)\left(\frac{\sqrt{\omega_{2}}}{\omega_{\rm{ln}}}\right).
\end{equation}
The parameter $\omega_{2}$ is the second moment of the normalized weight function, $\omega_{{\rm ln}}$ denotes the logarithmic phonon frequency and $\lambda$ is called the electron-phonon coupling constant. In the case of Calcium under the pressure at $200$ GPa the following results have been obtained: $\sqrt{\omega_{2}}=35.92$ $\rm{meV}$, $\omega_{{\rm ln}}=29.98$ meV and $\lambda=1.228$.

%%%%%%%%%%%%%%%%%%%%%%%%%%%%%%%%%%%%%%%%%%%%%%%%%%%%%%%%%%%%%%%%%%%%%%%%%%%%%%%%%%%%%%%%%%%%%%%%%%%
\section{SUMMARY}
%%%%%%%%%%%%%%%%%%%%%%%%%%%%%%%%%%%%%%%%%%%%%%%%%%%%%%%%%%%%%%%%%%%%%%%%%%%%%%%%%%%%%%%%%%%%%%%%%%%    

The thermodynamic parameters of the superconducting state in Calcium under the pressure at 200 GPa have been analyzed in the paper. The numerical calculations in the framework of the Eliashberg approach have been made. On the basis of the exact Eliashberg solutions the specific heats and the thermodynamic critical field have been determined.

For the wide range of the Coulomb pseudopotential values ($\mu^{\star}\in\left<0.1,03\right>$), it has been shown that the specific heat's jump at the critical temperature and the thermodynamic critical field near zero Kelvin strongly decrease with $\mu^{\star}$. 

The dimensionless ratios $r_{1}$ and $r_{2}$ significantly differ from the predictions based on the BCS model even for high values of the Coulomb pseudopotential. In particular, $r_{1}\in\left<2.64,1.97\right>$ and $r_{2}\in\left<0.140,0.157\right>$.

%%%%%%%%%%%%%%%%%%%%%%%%%%%%%%%%%%%%%%%%%%%%%%%%%%%%%%%%%%%%%%%%%%%%%%%%%%%%%%%%%%%%%%%%%%%%%%%%%%%
\begin{acknowledgments}
The authors wish to thank Prof. K. Dzili{\'n}ski for providing excellent working conditions and the financial support. We also thank Mr M.W. Jarosik and Mr D. Szcz{\c{e}}{\'s}niak for the productive scientific discussion, that improved the quality of the presented paper. All numerical calculations were based on the Eliashberg function sent to us by: Prof. W.E. Pickett and Dr Z.P. Yin for whom we are very thankful.
\end{acknowledgments}
%%%%%%%%%%%%%%%%%%%%%%%%%%%%%%%%%%%%%%%%%%%%%%%%%%%%%%%%%%%%%%%%%%%%%%%%%%%%%%%%%%%%%%%%%%%%%%%%%%%
%
%%%%%%%%%%%%%%%%%%%%%%%%%%%%%%%%%%%%%%%%%%%%%%%%%%%%%%%%%%%%%%%%%%%%%%%%%%%%%%%%%%%%%%%%%%%%%%%%%%%

%%%%%%%%%%%%%%%%%%%%%%%%%%%%%%%%%%%%%%%%%%%%%%%%%%%%%%%%%%%%%%%%%%%%%%%%%%%%%%%%%%%%%%%%%%%%%%%%%%%
%

\begin{thebibliography}{99}
%
%%%%%%%%%%%%%%%%%%%%%%%%%%%%%[1]
\bibitem{Schilling}
(a) J.S. Schilling, High Pressure Res. {\bf 26}, 145 (2006);\\
(b) J.S. Schilling, J.J. Hamlin, Journal of Physics: Conference Series {\bf 121}, 052006 (2008). 
%%%%%%%%%%%%%%%%%%%%%%%%%%%%%[2]
\bibitem{Deemyad}
(a) S. Deemyad, J.S. Schilling, Phys. Rev. Lett. {\bf 91}, 167001 (2003);\\
(b) M. Luders, M.A.L. Marques, A. Floris, G. Profeta, N.N. Lathiotakis, C. Franchini, A. Sanna, A. Continenza, S. Massidda, E.K.U. Gross, 
    Psik Newsletter, Scientific Highlight of the month, {\bf 76}, (2006);\\
(c) R. Szcz{\c{e}}{\`s}niak, M.W. Jarosik, D. Szcz{\c{e}}{\`s}niak, Physica B {\bf 405}, 4897 (2010).
%%%%%%%%%%%%%%%%%%%%%%%%%%%%%[3]
\bibitem{Hamlin}
(a) J.J. Hamlin, V.G. Tissen, J.S. Schilling, Physica C {\bf 451}, 82 (2007);\\
(b) Z.P. Yin, S.Y. Savrasov, W.E. Pickett, Phys. Rev. B {\bf 74}, 094519 (2006).
%%%%%%%%%%%%%%%%%%%%%%%%%%%%%[4]
\bibitem{Yabuuchi}
T. Yabuuchi, T. Matsuoka, Y. Nakamoto, K. Shimizu, J. Phys. Soc. Jpn. {\bf 75}, 083703 (2006).
%%%%%%%%%%%%%%%%%%%%%%%%%%%%%[5]
\bibitem{Okada}
S. Okada, K. Shimizu, T.C. Kobayashi, K. Amaya, S. Endo, J. Phys. Soc. Jpn. {\bf 65}, 1924 (1996).
%%%%%%%%%%%%%%%%%%%%%%%%%%%%%[6]
\bibitem{Yin}
Z.P. Yin, F. Gygi, W.E. Pickett, Phys. Rev. B {\bf 80}, 184515 (2009).
%%%%%%%%%%%%%%%%%%%%%%%%%%%%%[7]
\bibitem{Eliashberg}
For discussion of the Eliashberg equations (originally formulated by G.M. Eliashberg, Soviet.
Phys. JETP {\bf 11}, 696 (1960)) we refer to: P.B. Allen, B. Mitrovi{\'c}, in: Solid State
Physics: Advances in Research and Applications, edited by H. Ehrenreich, F. Seitz, D. Turnbull, (Academic, New York, 1982), Vol 37, p. 1.
%%%%%%%%%%%%%%%%%%%%%%%%%%%%%[8]
\bibitem{Marsiglio}
J.P. Carbotte, F. Marsiglio, in: The Physics of Superconductors, edited by K.H. Bennemann, J.B. Ketterson, (Springer, Berlin, 2003), Vol 1, p. 223.
%%%%%%%%%%%%%%%%%%%%%%%%%%%%%[9]
\bibitem{Carbotte}
J.P. Carbotte, Rev. Mod. Phys. {\bf 62}, 1027 (1990).
%%%%%%%%%%%%%%%%%%%%%%%%%%%%%[10]
\bibitem{McMillan}
W.L. McMillan, J.W. Rowell, Phys. Rev. Lett. {\bf 14}, 108 (1965).
%%%%%%%%%%%%%%%%%%%%%%%%%%%%%[11]
\bibitem{Bardeen}
J. Bardeen, M. Stephen, Phys. Rev. {\bf 136}, A1485 (1964).
%%%%%%%%%%%%%%%%%%%%%%%%%%%%%[12]
\bibitem{BCS}
(a) J. Bardeen, L.N. Cooper, J.R. Schrieffer, Phys. Rev. {\bf 106}, 162 (1957);\\ 
(b) J. Bardeen, L.N. Cooper, J.R. Schrieffer, Phys. Rev. {\bf 108}, 1175 (1957).
%%%%%%%%%%%%%%%%%%%%%%%%%%%%%[13]
\bibitem{SzczesniakDurajski}
(a) R. Szcz{\c{e}}{\`s}niak, A.P. Durajski, arXiv:1105.5337;\\
(b) R. Szcz{\c{e}}{\`s}niak, A.P. Durajski, M.W. Jarosik, arXiv:1105.5345.
%%%%%%%%%%%%%%%%%%%%%%%%%%%%%[14]
\bibitem{SzczesniakSzczesniak}
R. Szcz{\c{e}}{\`s}niak, D. Szcz{\c{e}}{\`s}niak, unpublished results.
%%%%%%%%%%%%%%%%%%%%%%%%%%%%%[15]
\bibitem{Szczesniak}
(a) R. Szcz{\c{e}}{\`s}niak, Acta Phys. Pol. A {\bf 109}, 179 (2006);\\
(b) R. Szcz{\c{e}}{\`s}niak, Solid State Commun. {\bf 138}, 347 (2006);\\
(c) R. Szcz{\c{e}}{\`s}niak, M.W. Jarosik, Solid State Commun. {\bf 149}, 2053 (2009).
%%%%%%%%%%%%%%%%%%%%%%%%%%%%%[16]
\bibitem{SzczesniakDurajski1}
R. Szcz{\c{e}}{\`s}niak, A.P. Durajski, unpublished results.
%
\end{thebibliography}
\end{document}